# Performance Evaluation of Enhanced Interior Gateway Routing Protocol in IPv6 Network

Kuwar Pratap Singh
Department of Computer Science and Engineering
Jaypee University of Information Technology, Waknaghat, Solan
173 234 INDIA

P. K. Gupta
Department of Computer Science and Engineering
Jaypee University of Information Technology, Waknaghat, Solan
173 234 INDIA

G. Singh
Department of Electronics and Communication Engineering
Jaypee University of Information Technology, Waknaghat, Solan
173 234 INDIA

## ABSTRACT
With the explosive growth in communication and network technologies, there is a great demand of IPv6 addressing scheme. However, the modern operating systems has option for this and with the development of IPv6 which removes the limitations imposed by IPv4 and provides the large number of address space. In this paper, authors have considered the Enhanced Interior Gateway Routing Protocol and presented a scenario for its performance evaluation in IPv6 networks and obtained results are highly considerable for the short distance of communication and don't represent any problem of performance degradation while sending or receiving the data.

## 1. INTRODUCTION
Recently, a lot of developments have been taken place in computer networks and now it is the necessity that the computer in one geographical location must communicate with another computer located at different geographical location. For doing so, though there are various available alternatives but almost all alternatives require use of Internet Protocol (IP) address to uniquely identify the machine over the network. If the two different computers are in the same location then they can communicate via Local Area Network (LAN), if not, then different available alternatives can be used for connectivity and communication over the networks and Wide Area Network (WAN) is one of them. To communicate over networks, numbers of devices are required and router is one of them and the routing protocol must be configured properly on each router. This routing protocol maintains the routing table which further maintains the shortest path of communication from source to destination. Commonly, this routing protocol is in IPv4 but nowadays IPv4 addresses are depleting so these protocols are now using IPv6 addresses. There are number of dynamic routing protocols (DRP) supported by IPv6 such as RIPng, EIGRP, OSPFv3, IS-IS, and Multiprotocol BGP etc.

In this paper, authors have considered the Enhanced Interior Gateway Routing Protocol (EIGRP) [1, 2, 3] for IPv6 as a routing protocol, which is a Cisco proprietary protocol and uses classless routing schemes based on diffusion update algorithm (DUAL) for calculating a shortest path from source to destination. It also supports unequal path load balancing that is the main reason of its fastest protocol. Its administrative distance is 90 that represent the trustworthiness of protocol. EIGRP selects fastest path based on delay, bandwidth, reliability, and load. For this, smaller size of packets is transmitted after every 5 seconds to see whether neighbour router is up or not. These packets are sent for various different reasons like:

a) For discovering neighbour,

b) For forming relation with neighbour,

c) For maintaining relation with neighbour.

DUAL algorithm used, includes Successor(S), Feasible Successor (FS), Feasibility Condition (FC), Feasible Distance (FD), and reported distance (RD) [4].

## 2. RELATED WORK
Narisetty and Balsu [5] have introduced the combination of IS-IS/RIP and EIGRP protocol. In their obtained experimental results they have shown that the combination of IS-IS/RIP protocol gives better throughput and end-to-end delay in comparison to EIGRP protocol. Whereas, the network convergence of EIGRP protocol is better than IS-IS/RIP protocol. Yehia et al. [6] have proposed the various routing protocols and evaluated them based on some performance metrics. This evaluation is performed theoretically and by using simulation. EIGRP gives the best conversion duration and was the first to converge. In [7] Sunjian and Fang, have introduced the OSPF protocol for IPv6 which is also referred as OSPFv3 and They first introduced the knowledge of IPv6 and then implemented the OSPF over IPv6. In [8] Wijaya has proposed OSPF and EIGRP protocol in IPv4 and IPv6 network and simulated these protocols on some network topology and obtained results represent that EIGRP is better than OSPF in various topologies. Here, packet sent in an IPv4 environment is smaller than the packet sents in an IPv6 environment. This is because in the IPv6 network, addressing is much larger than in IPv4. In [9] Thorenoor, has introduced the implementation decisions to be made when the choice is available between protocols that involve distance vector or link state or the combination of both. In this paper, it is shown that EIGRP compared to RIP and OSPF definitely performs better compared to RIP and OSPF in terms of network convergence activity, packet loss, throughput, and end-to-end delay. Jaafar et al. [10] have introduced the EIGRP protocol which is developed by Cisco and calculated the shortest path using Diffusion Update Algorithm (DUAL). DUAL does not create the routing loops while computing the shortest path. They have also developed a detailed simulation model of EIGRP and evaluated the EIGRP performance. In, the obtained results they have found that EIGRP converges faster than single TCP timeout in many cases. Kalyan et al. [11] have introduced the dynamic routing protocols and discussed about the issue of choosing an optimal algorithm and according to the network specification, optimal algorithm is selected. Sanjeev et al. [12] presented a comparative analysis of throughput, delay and queue length for the various congestion control algorithms RED, SFQ and REM using NS2 simulator. They also included the comparative analysis of loss rate having different bandwidth for these algorithms. In [13] Nazrul and Ullah designed three network models configured with OSPF, EIGRP to evaluate the OSPF and EIGRP's performance, and all the three topologies were simulated using the Optimized Network Engineering Tool (OPNET). In this case, the protocols and the combined use of them are also analyzed in terms of convergence time, jitter, end-to-end delay, throughput, and packet loss. The evaluation results show that, in general, the combined implementation of EIGRP and OSPF routing protocols in the network performs better than each one of them alone.

## 3. IPv6
IPv6 is a 128 bits or 16 bytes addressing scheme, which is represented by a series of eight 16 bits field separated by colons [14]. The format of IPv6 is x:x:x:x:x:x:x:x ,where x is 16 bits hexadecimal numbers with leading zeros in each x field are optional. Successive x fields with 0 can be represented as :: but only once, for example 2031:0000:0000:013f:0000:0000:0000:0001. Security in IPv6 is inherit, which provides authentication and encryption and has simple header format for higher processing. Here, Table 1 represents the common difference between IPv4 and IPv6.

**Table1: Comparison between IPv4 and IPv6**

| IP Versions | Deployment year | Address Size | Number of addresses available | Format of address | Prefix notation |
|---|---|---|---|---|---|
| IPv4 | 1981 | 32 | $2^{32}$=4,294,967,296 | Dotted decimal notation | 192.168.0.0/24 |
| IPv6 | 1999 | 128 | $2^{128}$=approximately $3.4 \times 10^{38}$ addresses | Hexadecimal notation | 2001:abcd:abcd::/48 |

*Configuration of IPv6 on router*

a) TO enable IPv6 Routing

Router (config) # ipv6 unicast-routing

b) Configure IPv6 address

Router (config-if)# ipv6 address <IPv6 Address>/<prefix>

## 4. EIGRP IN IPv6
This is an enhanced distance vector protocol, which relies on the Diffused Update Algorithm (DUAL) to calculate the shortest path to a destination within a network. EIGRP in IPv6 works in the same way as EIGRP in IPv4 but can be configured and managed separately.

### 4.1 Unchanged Features of EIGRP in IPv6
- It's Cisco proprietary.
- Uses Dual algorithm.
- Metric composite is same.
- Updates are multicast by using multicast address FF02::A.
- Authentication

- Link bandwidth percentage
- Split horizon
- Hello interval and hold time configuration.
- Address summarization
- Stub router
- Variance

### 4.2 Changed Feature of EIGRP in IPv6
The changed features of EIGRP in IPv6 are listed as follows [15]:

- Configuration of Interface- Here, interfaces can be configured directly with EIGRP for IPv6, without using the global IPv6 address. There is no network statement in EIGRP for IPv6.
- Must no shutdown the routing process- EIGRP for IPv6 has a shutdown feature. The routing process should be in "no shutdown" mode in order to start running.
- Router ID- The router ID needs to be configured for an EIGRP IPv6 protocol instance before it can run.
- Route filtering- IPv6 EIGRP performs route filtering using only the distribute-list prefix-list command. EIGRP in IPv6 does not support route maps.
- No concept of automatic route summarization- EIGRP in IPv4 uses automatic route summarization in classful network but in IPv6 there is no concept of classful networks hence there is no concept of automatic summarization.

## 5. Implementation of EIGRP in IPv6
Here, authors have used simulation strategy by using packet tracer software. In this paper, authors have proposed a network topology, which consists of two routers and each router is connected to personal computer through switch. In this topology as shown in Fig 1, two different networks are used where one network consists of PC0, PC1, SWITCH0, ROUTER0 and another network consists of PC2, SWITCH1, and ROUTER1. PC0 consist of IPV6 address 2001:11:11:11::10/64, PC1 consist of IPv6 address 2001:11:11:11::11/64 and PC2 consist of IPv6 address 2012:13:13:13::20/64. If PC0 wants to communicate with PC1 then there is no need of router as they both are the part of same network and can communicate through intermediate device switch, but if PC0 or PC1 wants to communicate with PC2 then there is need of router as they both are the part of different networks. Two interfaces, Ethernet and serial port are required on the router to enable this communication. Let Router 0 consist of two interfaces known as Ethernet 0/0 and Serial 0/0/1. Ethernet 0/0 consist IPv6 address 2001:11:11:11::1/64 and serial0/0/1 consist IPv6 address 2010:AB8::1/64. Router 1 consist of two interface known as Ethernet 0/0 and Serial 0/0/0. Here, Ethernet 0/0 consist IPv6 address 2012:13:13:13::1/64 and serial 0/0/0 consist IPv6 address 2010:AB8::2/64.

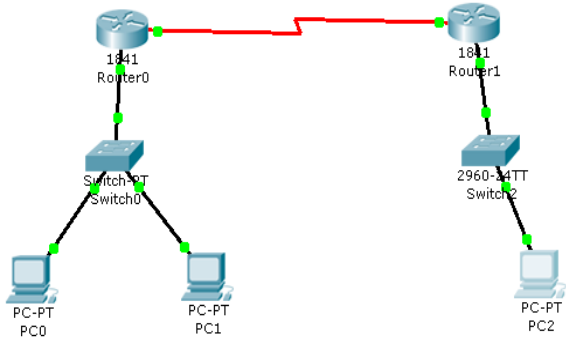

**Fig1: Implemented topology for performance evaluation.**

## 5.1 Main configuration command

These are the commands which are used to configure the IPv6 EIGRP protocol on the router 0 and router 1, which can learn the shortest path and deliver the packet between different network through a corresponding shortest path. IPv6 unicast-routing command is used for enabling IPv6 forwarding, which will be the first command as after enabling IPv6 another command can be executed. IPv6 unicast-routing command is configured on the router in its configuration mode and no shut command is mandatory as to start the routing process. The various router configurations are shown as follows:

### 5.1.1  Router 0 configuration

hostname Router0
!
ipv6 unicast-routing//ipv6 forwarding is enabled//
!
interface Fastethernet0/0
ipv6 address 2001:11:11:11::1/64
//configure IPv6 address for interface Fastethernet0/0//
no shutdown// Enables no shut mode so the routing process can start running//
ipv6 enable// IPv6 processing on the fastethernet 0/0 interface//
ipv6 eigrp 10// Enables the EIGRP for IPv6 process on the fastethernet 0/0 interface
interface Serial0/0/1
ipv6 address FE80::1 link-local
ipv6 address 2010:AB8::1/64
// configure IPv6 address for interface Serial0/0/1//

no shutdown
ipv6 enable// Enables IPv6 processing on the Serial0/0/1 interface//
ipv6 eigrp 10// Enables the EIGRP for IPv6 process on the Serial0/0/1 interface//
clock rate 2000000
!
ipv6 router eigrp 10//eigrp ipv6 routing protocol is enabled//
 router-id 2.2.2.2//Router Id is defined//
 no shutdown// Enables no shut mode so the routing process can start running//
!
end

### 5.2.2 Router 1 configuration

hostname Router0
!
ipv6 unicast-routing//ipv6 forwarding is enabled//
!
interface Fastethernet0/0
ipv6 address 2012:13:13:13::1/64
//configure IPv6 address for interface Fastethernet0/0//
no shutdown// Enables no shut mode so the routing process can start running//
ipv6 enable// IPv6 processing on the fastethernet 0/0 interface//
ipv6 eigrp 10// Enables the EIGRP for IPv6 process on the fastethernet 0/0 interface
interface Serial0/0/1
ipv6 address FE80::2 link-local
ipv6 address 2010:AB8::2/64
// configure IPv6 address for interface Serial0/0/1//

no shutdown
ipv6 enable// Enables IPv6 processing on the Serial0/0/1 interface//
ipv6 eigrp 10// Enables the EIGRP for IPv6 process on the Serial0/0/1 interface//
clock rate 2000000
!
ipv6 router eigrp 10//eigrp ipv6 routing protocol is enabled//
 router-id 1.1.1.1//Router Id is defined//
 no shutdown// Enables no shut mode so the routing process can start running//

!
end

## 6. EXPERIMENTS RESULTS

This section represents the various obtained results for both the router 0 and router 1 in the form of routing table, routing protocol information, IPv6 EIGRP neighbor information, and interface information.

### 6.1  Routing table

Routing table of router 0 and router 1 as shown in Table 2 and Table 3 maintains the shortest path to other routers. The letter C represents that the routing is directly connected to the router 0. The letter D represents that the routing is local and is done by using EIGRP. The following command shows IPv6-specific EIGRP routes:

**Routing table statistics:**

IPv6 Routing Table - 6 entries

Codes: C - Connected, L - Local, S - Static, R - RIP, B - BGP

U - Per-user Static route, M - MIPv6

I1 - ISIS L1, I2 - ISIS L2, IA - ISIS interarea, IS - ISIS summary

O - OSPF intra, OI - OSPF inter, OE1 - OSPF ext 1, OE2 - OSPF ext 2

ON1 - OSPF NSSA ext 1, ON2 - OSPF NSSA ext 2

D - EIGRP, EX - EIGRP external

**Command:** router0#sh ipv6 route

**Table 2: IPv6-routing table for router 0.**

| Codes | IP Address | LINK |
|---|---|---|
| C(Directly connected) | 2001:11:11:11::/64 [0/0] | via::, FastEthernet0/0 |
| L(Locally connected) | 2001:11:11:11::1/128 [0/0] | via::, FastEthernet0/0 |
| C(Directly connected) | 2010:AB8::/64 [0/0] | via ::, Serial0/0/1 |
| L(Locally connected) | 2010:AB8::1/128 [0/0] | via ::, Serial0/0/1 |
| D(Route learned by EIGRP) | 2012:13:13:13::/64 [90/2172416] | via FE80::2, Serial0/0/1 |
| L(Locally connected) | FF00::/8 [0/0] | via ::, Null0 |

**Command:** router1#sh ipv6 route

**Table 3: IPv6-routing table for router 1.**

| Codes | IP Address | LINK |
|---|---|---|
| D(Route learned by EIGRP) | 2001:11:11:11::/64 [90/2172416] | via FE80::1, Serial0/0/0 |
| C(Directly connected) | 2010:AB8::/64 [0/0] | via ::, Serial0/0/0 |
| L(Locally connected) | 2010:AB8::2/128 [0/0] | via ::, Serial0/0/0 |
| C(Directly connected) | 2012:13:13:13::/64 [0/0] | via ::, FastEthernet0/0 |
| L(Locally connected) | 2012:13:13:13::1/128 [0/0] | via ::, FastEthernet0/0 |
| L(Locally connected) | FF00::/8 [0/0] | via ::, Null0 |

## 6.2 Routing protocol information

The routing protocol information shows that EIGRP IPV6 routing protocol has been established and assigned process ID is 10. EIGRP is enabled on interface fast Ethernet0/0, serial0/0/0 and serial0/0/1.Maximum hop count supported by EIGRP is 100.

**Command:** router0#sh ipv6 protocol

IPv6 Routing Protocol is "connected"
IPv6 Routing Protocol is "static
IPv6 Routing Protocol is "eigrp 10 "
  EIGRP metric weight K1=1, K2=0, K3=1, K4=0, K5=0
  EIGRP maximum hopcount 100
  EIGRP maximum metric variance 1
  Interfaces:
    FastEthernet0/0
    Serial0/0/0
    Serial0/0/1
  Redistributing: eigrp 10
  Maximum path: 16
  Distance: internal 90 external 170

**Command:** router1#sh ipv6 protocol

IPv6 Routing Protocol is "connected"
IPv6 Routing Protocol is "static
IPv6 Routing Protocol is "eigrp 10 "
  EIGRP metric weight K1=1, K2=0, K3=1, K4=0, K5=0
  EIGRP maximum hopcount 100
  EIGRP maximum metric variance 1
  Interfaces:
    FastEthernet0/0
    Serial0/0/0
    Serial0/0/1
  Redistributing: eigrp 10
  Maximum path: 16
  Distance: internal 90 external 170

## 6.3 IPv6 EIGRP neighbor information

The obtained results also display the neighbors discovered by the EIGRP IPv6. It shows the link for local addresses of router0 and router1. These results are listed in Table 4 and Table 5.

**Command:** router0#sh ipv6 eigrp neighbour

**Table 4: IPv6-EIGRP neighbors for process 10(router 0)**

| H | 0 |
|---|---|
| Address | FE80::2 |
| Interface (Sec) | Se0/0/1 |
| Hold (ms) | 12 |
| Uptime Seq Cnt | 00:21:14 |
| SRTT Num | 40 |
| RTO | 1000 |
| Q | 0 |
| Seq Num | 3 |

**Command:** router1#sh ipv6 eigrp neighbour

**Table 5: IPv6-EIGRP neighbors for process 10 (router 1)**

| H | 0 |
|---|---|
| Address | FE80::1 |
| Interface (Sec) | Se0/0/0 |
| Hold (ms) | 14 |
| Uptime Seq Cnt | 00:22:23 |
| SRTT Num | 40 |
| RTO | 1000 |
| Q | 0 |
| Seq Num | 3 |

## 6.4 IPv6 EIGRP Interface information

The EIGRP IPv6 interface information displays information about the interfaces configured for EIGRP in IPv6 for both the routers 0 and 1 and displayed in Table 6 and Table 7.

**Command:** router0#show ipv6 eigrp interfaces

**Table 6: IPv6-EIGRP interfaces for process 10 using router 0.**

| Interface | Peers Routes | Xmit Queue Pending Un/Reliable | Mean SRTT | PacingTime Un/Reliable | Multicast Flow Timer | Pending routes |
|---|---|---|---|---|---|---|
| Fa0/0 Se0/0/1 | 0 1 | 0/0 0/0 | 1236 1236 | 0/0 0 | 0/10 0/10 | 0 0 |

The output of this command displays information about interfaces configured for EIGRP in IPv6. For router 0 two interfaces Fa0/0 and Se0/0/1 are configured and the command determines which interface of EIGRP is active.

**Command:** router1#show ipv6 eigrp interfaces

**Table 7: IPv6-EIGRP interfaces for process 10 using router 1.**

| Interface | Peers Routes | Xmit Queue Pending Un/Reliable | Mean SRTT | Pacing Time Un/Reliable | Multicast Flow Timer | Pending routes |
|---|---|---|---|---|---|---|
| Se0/0/0 Fa0/0 | 1 0 | 0/0 0/0 | 1236 1236 | 0/10 0/10 | 0 0 | 0 0 |

The output of this command Displays information about interfaces configured for EIGRP in IPv6. For Router 1 two interfaces Fa0/0 and Se0/0/0 are configured and the command determines which interface of EIGRP is active.

## 6.5 EIGRP IPv6 Topology information

To display output of the EIGRP IPv6 topology table, authors used the EIGRP IPv6 topology command in privileged EXEC mode. The following EIGRP IPv6 topology command can be used to determine diffusion update algorithm (DUAL) states and to debug possible DUAL problems. Here, obtained results are displayed in Table 8 and Table 9.

**Codes:** P - Passive, A - Active, U - Update, Q - Query, R - Reply, r - Reply status

**Command:** router0#sh ipv6 EIGRP topology

**Table 8: IPv6-EIGRP Topology Table for AS 10/ID(2.2.2.2)**

| Codes | IP Address | Successor | FD | Link |
|---|---|---|---|---|
| P | 2001:11:11:11::/64 | 1 | FD is 28160 | Via Connected, FastEthernet0/0 |
| P | 2010:AB8::/64, | 1 | FD is 2169856 | Via Connected, Serial0/0/1 |
| P | 2012:13:13:13::/64 | 1 | FD is 2172416 | Via FE80::2 (2172416/28160), Serial0/0/1 |

**Command:** router1#sh ipv6 EIGRP topology

**Table 9: IPv6-EIGRP Topology Table for AS 10/ID(1.1.1.1)**

| Codes | IP Address | Successor | FD | Link |
|---|---|---|---|---|
| P | 2012:13:13:13::/64 | 1 | FD is 28160 | Via Connected, FastEthernet0/0 |
| P | 2010:AB8::/64, | 1 | FD is 2169856 | Via Connected, Serial0/0/0 |
| P | 2001:11:11:11::/64 | 1 | FD is 2172416 | Via FE80::2 (2172416/28160), Serial0/0/1 |

## 7. CONCLUSION

Routing is the process of moving packets from one network to the other network. Basically, the routing involves two activities: determining best path and forwarding packets through this path. These two activities of routing vary on the basis of selection of routing protocol. However, various routing protocol selects different selection process for the best path. Routing protocols such as RIP, OSPF, IS-IS, and EIGRP directly impacts on internet efficiency. So these routing protocols should use IPv6 address as the IPv4 addresses are depleted day-by-day. EIGRP protocol has better routing capabilities hence EIGRP IPv6 will be mainly used in the future.